\documentclass[11pt]{article}
\usepackage{amsmath}
\usepackage{amsfonts}
\usepackage{amssymb}
\usepackage{epsfig}
\usepackage{times}
\newcommand{\be}{\begin{equation}}
\newcommand{\ee}{\end{equation}}
\newcommand{\ba}{\begin{array}}
\newcommand{\ea}{\end{array}}
\newcommand{\bea}{\begin{eqnarray}}
\newcommand{\eea}{\end{eqnarray}}

\newcommand{\rar}{\rightarrow}
\newcommand{\p}{\partial}
\newcommand{\ol}{\overline}
\newcommand{\ti}{\tilde}

\newcommand{\bs}{\boldsymbol}

\renewcommand{\l}{\newline\null}
\def\figskip{\vskip .5cm plus 3mm minus 2mm}
\begin{document}
\begin{titlepage}
December 2000\hfill PAR-LPTHE 00/48
\vskip 4.5cm
{\baselineskip 17pt
\begin{center}
{\bf GAUGE BOSONS IN AN $\mathbf {SU(2)_L \times
SU(2)_R} \times G_{lept}$ ELECTROWEAK MODEL}
\end{center}
}
\vskip .5cm
\centerline{B. Machet
     \footnote[1]{Member of `Centre National de la Recherche Scientifique'}
     \footnote[2]{E-mail: machet@lpthe.jussieu.fr}
     }
\vskip 5mm
\centerline{{\em Laboratoire de Physique Th\'eorique et Hautes \'Energies}
     \footnote[3]{LPTHE tour 16\,/\,1$^{er}\!$ \'etage,
          Universit\'e P. et M. Curie, BP 126, 4 place Jussieu,
          F-75252 Paris Cedex 05 (France)}
}
\centerline{\em Universit\'es Pierre et Marie Curie (Paris 6) et Denis
Diderot (Paris 7)}
\centerline{\em Unit\'e associ\'ee au CNRS UMR 7589}
\vskip 1cm
{\bf Abstract:}  By considering its generalization to composite
 $J=0$ mesons proposed in \cite{Machet1},
I show how and why a chiral extension of the Glashow-Salam-Weinberg standard
model of electroweak interactions calls, there, for right-handed charged
$W^\pm_{\mu R}$'s coupled with $g_R = e/\cos\theta_W$, and the masses of
which are related to the ones of the left-handed $W^\pm_{\mu L}$'s through the
relation $M_{W_{L}}^2 + M_{W_{R}}^2  = M_Z^2$.
The mesonic sector, having vanishing baryonic and leptonic number, is neutral
with respect to the corresponding  $U(1)_{\mathbb I}$ symmetries, making the
natural chiral gauge group to be $SU(2)_L \times SU(2)_R$, blind to the
presence of extra $Z'_\mu$'s. 
The $W^\pm_{\mu R}$ gauge bosons
cannot have been detected in hadronic colliders and can be very elusive in
electroweak processes involving, in particular, pseudoscalar mesons.
Present data select one among two possible extensions for which, in the
right sector: -- a specific breaking of universality
occurs between families of quarks, which belong to inequivalent
representations of $SU(2)_R$; -- the mixing angle is a free parameter,
constrained to be smaller than the Cabibbo angle by the box diagrams
controlling the $K_L-K_S$ mass difference; this also minimizes
contributions to $\mu \rightarrow e \gamma$.
The relation $g_L^2/M^2_{W^\pm_L}= g_R^2/M^2_{W^\pm_R}$ implements
left-right symmetry for low energy charged effective weak interactions.
For the sake of simplicity, this study is performed for two generations only.
\smallskip

\vfill
\null\hfil\epsffile{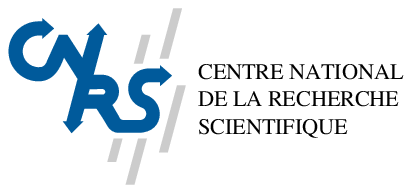}
\end{titlepage}
%
\section{Introduction.}
\label{section:introduction}
%
The Glashow-Salam-Weinberg standard model of electroweak interactions
accounts for parity violation by enforcing it from the start at the level of
the gauge group of symmetry; one would rather like the asymmetry
observed in nature to arise from dynamical considerations. This was one of
the goals of left-right symmetric models, in particular the ones
based on the gauge group $SU(2)_L \times SU(2)_R \times U(1)_{B-L}$
\cite{LangackerSankar}.

We too shall be concerned here with finding a chiral extension to
the standard electroweak model; more precisely, we shall extend to a
chiral form the already achieved extension of the standard model to $J=0$
states transforming like fermion-antifermion composite fields
\footnote{It
includes from the start the chiral and electroweak properties of quarks
(which are no longer fields of the Lagrangian).}
.

The main step is to identify the weak hypercharge generator $\mathbb Y$
with the right-handed ${\mathbb T}^3_R$ generator; this is possible for
mesons because the extra purely vectorial $U(1)_{\mathbb I}$ part of
$U(1)_\mathbb Y$
does not act on composite fermion-antifermion fields;  essential at the
fermionic (leptonic) level, it is of no relevance as far as the spectrum of
the gauge fields, entirely determined at the hadronic level by the vacuum
expectation values of composite mesonic fields, is concerned.
Hence, the left-right symmetric extension that we are led to consider
here involves only the group  $SU(2)_L \times SU(2)_R$, with left and right
coupling constants being respectively $e/\sin\theta_W$ and
$e/\cos\theta_W$, as suggested by the Gell-Mann-Nishijima relation written
in chiral form; the left-right symmetry \cite{SenjanovicMohapatra},
absent at the level of the Lagrangian, is however implemented at
low energy for effective charged weak interactions.

Both the left- and right- $SU(2)$'s have the same generator ${\mathbb T}^3$:
it is a
constraint from which to build the sought for $SU(2)_R$ group; 
the diagonal $U(1)$ with generator ${\mathbb T}^3$ is the only one which can
be left unbroken in the spontaneous breaking of the electroweak symmetry;
it corresponds  to the electric charge in the mesonic sector.

The electroweak symmetry breaking is triggered by the non-vanishing
vacuum expectation value of a scalar(s) meson(s) (Higgs boson), which
behaves itself like a quark-antiquark composite; all $J=0$ composite
representations depend on the mixing angles and are, in the case of
$SU(2)_L$,
isomorphic to the scalar quadruplet of the standard model; their $SU(2)_R$
counterparts are easily determined but, in general, none is a rep. of both
$SU(2)_L$ and $SU(2)_R$
 such that the $SU(2)_L \times SU(2)_R$ symmetry breaking scalar
potential has to make use of appropriate combinations of them.
No Higgs multiplet with non-vanishing lepton ($B-L$) number
\cite{MohapatraSenjanovic} is introduced.

We are concerned in this work with the spectrum of the gauge fields,  
constrained by the observed properties of the $Z_\mu$ and $W_{\mu L}^\pm$.
The question of an extra $Z'_\mu$ does not arise for $SU(2)_L \times SU(2)_R$.
The charged sector is enlarged by  two $W_{\mu R}^\pm$ with masses
$M_{W_R}= M_Z \sin\theta_W = M_{W_L}\tan\theta_W \approx 43\,GeV$.

There happens to be two possible right-handed $SU(2)_R$ groups; this is linked
to the fact that the $N$-vector of fermions, which lies in the fundamental
representation of the diagonal subgroup of the chiral $U(N)_L \times
U(N)_R$, is reduceable with respect to the electroweak $SU(2)$'s and split
into $N/2$ doublets, one for each  family, which can
belong to inequivalent representations.

In each of the two possible extensions there occurs, in the right sector,
 the equivalent
$\varphi$ of the Cabibbo angle $\theta_c$  of the left-handed sector;
for the first type of $SU(2)_R$ 
it is constrained to be $\varphi = \theta_c + \pi/2$, while it is
left free in the second type; in this last case, the couplings
of the new gauge fields to mesons become $\varphi$-dependent;
we shall see that it is the one favoured by data for pseudoscalar
mesons.
In general, none of the relations between left and right mixing angles
commonly considered in the literature \cite{LangackerSankar} appears to be
realized in this approach: the right angles are
free parameters to be determined by experiment like their left
counterparts.

Light $W^\pm_{\mu R}$ like mentioned above have been unsought for, and we
show in particular that they cannot have been detected in hadronic
colliders, which started there investigations at a higher threshold.

They can eventually yield faint signals in electroweak processes,
and we investigate the special decay $D^+ \rar \pi^0 K^+ \pi^0$, seemingly
undetected; explaining this absence by a cancellation between
$W^\pm_{\mu L}$ and $W^\pm_{\mu R}$  discards the first
possible chiral extension and favors the one where, in the associated quark
picture, the two families belong to inequivalent representations of $SU(2)_R$.
Then, we show that the influence of $W^\pm_{\mu R}$ on the computation of
the $K_L-K_S$ mass difference at the quark level can be made negligeable when
the corresponding mixing angle is very small, without advocating for very
massive $W^\pm_{\mu R}$ \cite{BeallBanderSoni} .

Small enough right mixing angles also provides any desired suppression 
of unwanted contributions to $\mu \rar e\gamma$.

This study does not pretend to be exhaustive nor to provide a universal
chiral extension of the standard electroweak model; in particular, as
stressed again in the conclusion, it can hardly be conceived that dealing with
the leptonic sector does not require additional gauge field(s) and a more
complex group structure. Nevertheless, a chiral extension of the
standard model should accommodate the results presented here in the sector of
composite $J=0$ mesons.

We deal with $N/2 = 2$ generations of quarks.

\section{$\mathbf{SU(2)_L \times U(1)}$ as a truncated $\mathbf{SU(2)_L
\times SU(2)_R}$}

While an abelian $U(1)$ group can be chirally non-diagonal,
a non-abelian group can only be right-handed, left-handed or diagonal;
our goal being to ``extend'' the hypercharge $U(1)_{\mathbb Y}$ group of the
standard model to a non-abelian structure,  a special
attention to handedness is due.

Any $SU(2)$ (and $U(1)$) group can be considered as  a subgroup of
chiral $U(N)_L \times U(N)_R$ for $N$ even, and its generators taken as
$N \times N$ matrices.

Accordingly, we rewrite the Gell-Mann-Nishijima relation in its chiral form
\cite{Machet1}
\begin{equation}
({\mathbb Y}_L,{\mathbb Y}_R) = ({\mathbb Q}_L, {\mathbb Q}_R) - {\mathbb
T}^3_L,
\label{eq:GMN}
\end{equation}
where, in the hadronic sector
\begin{equation}
{\mathbb Q} = \left( \begin{array}{rrr}
                  \frac{2}{3}{\mathbb I}_2 & \vline &    \cr
                                          \hline
                 & \vline & -\frac{1}{3}{\mathbb I}_2
                     \end{array}\right),
\quad
{\mathbb T}^3 = \frac{1}{2}\left(\begin{array}{ccc}
             {\mathbb I}_2   &  \vline   &   \cr
                    \hline
                  &     \vline   &   -{\mathbb I}_2 \end{array}\right),
\label{eq:QT3}
\end{equation}
${\mathbb I}_2$ being the $2 \times 2$ identity matrix.

(\ref{eq:GMN}) also rewrites
\begin{equation}
{\mathbb Y}_{hadr} = \frac{1}{6} {\mathbb I}_4 + {\mathbb T}^3_R
\label{eq:GMN2}
\end{equation}
where ${\mathbb I}_4$ is the $4\times 4$ identity matrix
\footnote{In general, ${\mathbb Y} = \alpha {\mathbb I} + {\mathbb T}^3_R$,
with $\alpha = 1/6$ for hadrons and $\alpha = -1/2$ for leptons; in the
framework of a chiral theory where right-handed leptons are doublets of
$SU(2)_R$, $\alpha$ is interpreted as $(B-L)$ \cite{MarshakMohapatra}.}
.

 For any $4 \times 4$
matrix $\mathbb M$, a composite state $\ol\Psi (\gamma_5) {\mathbb M} \Psi$
is left invariant by the action of ${\mathbb I}_4$ since it is modified
\cite{Machet1} by
the commutator $\ol\Psi (\gamma_5)[{\mathbb I}_4,{\mathbb M}]\Psi$ when $\Psi$
and $\ol\Psi$ are acted upon.
Hence, when dealing with composite $J=0$ mesons (considered to be both the
fields of the Lagrangian and the asymptotic states of the theory)
the weak hypercharge  coincides with the right-handed generator
${\mathbb Y}_{hadr} \equiv {\mathbb T}^3_R$. It is the starting point for
building the looked for right handed $SU(2)_R$
\footnote{(\ref{eq:GMN2}) also writes ${\mathbb Q} =  \frac{1}{6} {\mathbb
I}_4 + {\mathbb T}^3$, which shows that, in the mesonic sector, for the
reasons just explained, the
electric charge generator $\mathbb Q$ coincides with ${\mathbb T}^3$.}
.

\subsection{Two types of $\mathbf{SU(2)}$}

According to the remarks above,
we look for the possible $SU(2)_R$ groups, with generators taken as $N
\times N \equiv 4 \times 4$ matrices, among which ${\mathbb T}^3$ is given in
(\ref{eq:QT3}).

Let ${\mathbb D}_1$ and ${\mathbb D}_2$ be the two $2 \times 2$ matrices
\begin{equation}
{\mathbb D}_1 = {\mathbb I}_2 = \left( \begin{array}{rr} 
                                           1 &     \cr
                                             &  1  \end{array} \right),
\quad
{\mathbb D}_2 = \left( \begin{array}{rr}    1 &    \cr
                                              & -1 \end{array} \right)
\end{equation}
and  ${\cal R}(\varphi)$ the rotation matrix
\begin{equation}
{\cal R}(\varphi) = \left( \begin{array}{rr} \cos\varphi   &   \sin\varphi   \cr
                                         -\sin\varphi  &   \cos\varphi
                  \end{array}\right),
\end{equation}
to which we associate the $4 \times 4$ matrix
\begin{equation}
{\mathbb R}(\varphi) = \left( \begin{array}{rrr}
                         {\mathbb I}_2 & \vline &      \cr
                                 \hline 
                                    &\vline   & {\cal R}(\varphi)
                   \end{array} \right).
\end{equation}
The first $SU(2)$, ${\cal G}_1$ has generators 
\bea
{\mathbb T}^3_1 &=& \frac{1}{2}
    {\mathbb R}^\dagger(\varphi)\left( \begin{array}{rrr}
       {\mathbb I}_2 & \vline  &     \cr
               \hline
            & \vline  & -{\mathbb I}_2\end{array}\right)
       {\mathbb R}(\varphi)
 = \frac{1}{2} \left( \begin{array}{rrr}
       {\mathbb I}_2 & \vline  &     \cr
               \hline
                  & \vline  & -{\mathbb I}_2\end{array}\right)
             \equiv {\mathbb T}^3  ,\cr
{\mathbb T}^+_1(\varphi) &=& {\mathbb R}^\dagger (\varphi)
\left(\begin{array}{rrr}
                 & \vline  & {\mathbb D}_1    \cr
                       \hline
                 &  \vline  &      \end{array} \right) {\mathbb R}(\varphi)
= \left( \begin{array}{rrr}  & \vline &{\cal R}(\varphi)   \cr
                        \hline
                             & \vline    & \end{array}\right),\cr
{\mathbb T}^-_1(\varphi) &=& {\mathbb R}^\dagger (\varphi) \left(
\begin{array}{rrr}
                 & \vline  &     \cr
                       \hline
    {\mathbb D}_1    &  \vline  &      \end{array} \right) {\mathbb
R}(\varphi)
= \left( \begin{array}{rrr}  & \vline &   \cr
                        \hline
            {\cal R}^\dagger (\varphi)     & \vline    & \end{array}\right).
\label{eq:G1}
\eea
and the second, ${\cal G}_2$, has generators

\vbox{
\bea
{\mathbb T}^3_2  &\equiv& {\mathbb T}^3_1 \equiv {\mathbb T}^3, \cr
{\mathbb T}^+_2(\varphi) &=&{\mathbb R}^\dagger(\varphi)\left(
\begin{array}{rrr}
                 & \vline   & {\mathbb D}_2    \cr
                       \hline
                 &  \vline  &      \end{array} \right) {\mathbb R}(\varphi)
= \left( \begin{array}{rrrr}  & \vline & \cos\varphi & \sin\varphi   \cr
                              & \vline & \sin\varphi & -\cos\varphi  \cr
                        \hline
                             & \vline  & & \end{array}\right),\cr
{\mathbb T}^-_2(\varphi) &=& {\mathbb R}^\dagger (\varphi) \left(
\begin{array}{rrr}
                 & \vline  &     \cr
                       \hline
    {\mathbb D}_2    &  \vline  &      \end{array} \right) {\mathbb
R}(\varphi)
= \left( \begin{array}{rrrr}  & & \vline &   \cr
                        \hline
             \cos\varphi & \sin\varphi & \vline &  \cr
            \sin\varphi & -\cos\varphi  & \vline    & \end{array}\right).
\label{eq:G2}
\eea
}

Transforming $\varphi$ into $\varphi + \pi/2$ is equivalent to changing, in
${\cal G}_1$, ${\mathbb D}_1$ into
\begin{equation}
{\mathbb D}_4 = \left( \begin{array}{rr}  & 1 \cr
                                        -1 & \end{array}\right)
\end{equation}
and transforming $\varphi$ into $\varphi + \pi/2$ in ${\cal G}_2$ equivalent
to going from ${\mathbb D}_2$ to
\begin{equation}
{\mathbb D}_3 = \left( \begin{array}{rr}    &  1  \cr
                                          1 &  \end{array}\right).
\end{equation}
The Glashow-Salam-Weinberg model uses ${\cal G}_1$ as the $SU(2)_L$ group
acting on the $4$-vector of quarks
\begin{equation}
\Psi = \left( \begin{array}{c} u\cr c\cr d\cr s \end{array}\right);
\label{eq:psi}
\end{equation}
$\Psi$, reduceable with respect to $SU(2)$, lies in the fundamental
representation of
the diagonal $U(4)$ subgroup of the chiral group $U(4)_L \times U(4)_R$;
${\cal R}(\varphi = \theta_c)$ is then the Cabibbo matrix $\mathbb C$ .

The eight $J=0$ composite representations of $SU(2)_L$ transforming like
quark-antiquark operators are built according to \cite{Machet1};

\vbox{
\bea & &\Phi_L({\mathbb D},\theta_c)= [{\mathbb M}\,^0_L, {\mathbb M}^3_L,
   {\mathbb M}^+_L, {\mathbb M}^-_L]({\mathbb D},\theta_c)\cr & &\ \cr
&=& {\mathbb R}^\dagger(\theta_c)
\left[  \frac{1}{\sqrt{2}}\left(\begin{array}{ccc}
      {\mathbb D} & \vline & 0\\
          \hline  0 & \vline & {\mathbb D}
                   \end{array}\right),
\frac{i}{\sqrt{2}} \left(\begin{array}{ccc}  {\mathbb D} & \vline & 0\\ 
\hline    0 & \vline & -{\mathbb D}
\end{array}\right),
i\left(\begin{array}{ccc} 0 & \vline & {\mathbb D}\\
    \hline   
 0 & \vline & 0           \end{array}\right),
 i\left(\begin{array}{ccc}  0 & \vline & 0\\ 
 \hline   {\mathbb D} & \vline & 0  \end{array}\right)
  \right]
{\mathbb R}(\theta_c)\cr
& & \cr
& & =\left[  \frac{1}{\sqrt{2}}\left(\begin{array}{ccc}
      {\mathbb D} & \vline & 0\\
          \hline  0 & \vline & {\mathbb C}^\dagger\,{\mathbb
D}\,{\mathbb C}
                   \end{array}\right),
\frac{i}{\sqrt{2}} \left(\begin{array}{ccc}  {\mathbb D} & \vline & 0\\ 
\hline    0 & \vline & -{\mathbb C}^\dagger\,{\mathbb
D}\,{\mathbb C}
\end{array}\right),
i\left(\begin{array}{ccc} 0 & \vline & {\mathbb D}\,{\mathbb C}\\
    \hline   
 0 & \vline & 0           \end{array}\right),
 i\left(\begin{array}{ccc}  0 & \vline & 0\\ 
 \hline    {\mathbb C}^\dagger\,{\mathbb D} & \vline & 0  \end{array}\right)
  \right],\cr
&&
\label{eq:repsL}
\eea
}
with ${\mathbb D} \in \{ {\mathbb D}_1 \cdots {\mathbb D}_4 \}$, and split
into the two types $({\mathbb M}^0, \vec {\mathbb M}) = ({\mathbb S}^0,
\vec{\mathbb P})$ and $({\mathbb M}^0, \vec {\mathbb M}) = ({\mathbb P}^0,
\vec{\mathbb S})$ where $\mathbb S$ denotes a scalar and $\mathbb P$ a
pseudoscalar; their laws of transformations are given in \cite{Machet1}.
They depend on the mixing angles and can be decomposed into two doublets
$(2)$ and $(\ol 2)$ of
$SU(2)_L$, or one singlet plus one triplet of the diagonal custodial
$SU(2)$. Each real quadruplet (\ref{eq:repsL}) is isomorphic to the 
complex scalar doublet of the standard model.

The meson field, of dimension [{\it mass}] attached to a given matrix
is obtained \cite{Machet1} by
sandwiching it between $\ol\Psi$ and $\Psi$, after eventually adding a
$\gamma_5$ matrix for pseudoscalar mesons, and introducing
an appropriate normalization factor \cite{Machet1}\cite{Machet2};
there is in particular a one-to-one correspondence between the quark content
of a meson and its matricial expression.

We shall consider for $SU(2)_R$ the two possibilities:\l
- $SU(2)_R = {\cal G}_1(\varphi)$, that we call the ``replica'' case because
  for $\varphi =\theta_c$, $SU(2)_R$ is the exact replica of the
standard $SU(2)_L$; in particular, when the mixing angle is turned off, the
two families of quarks doublets
\begin{equation}
\left( \begin{array}{c} u \cr d \end{array}\right) \quad \text{and} \quad
\left( \begin{array}{c} c \cr s \end{array}\right)
\end{equation}
making up the reduceable quadruplet $\Psi$ are acted upon in a similar way
by $SU(2)$;\l
- $SU(2)_R = {\cal G}_2(\varphi)$, where the two families belong to
 inequivalent representations of $SU(2)_R$, thus exhibiting in this sector
a specific breaking of universality; we call it the ``inverted'' case
to remind that  a ``$-$'' sign occurs relatively to the first family
when ${\mathbb T}^\pm_{2R}$ act on the second family.

We shall study them successively;  from the
fact that $\{{\mathbb D}_1,{\mathbb D}_2,{\mathbb D}_3,{\mathbb D}_4\}$ form a
complete set for real $2 \times 2$ matrices, one has exhausted, for the
case of two generations,
the possible extensions of  the standard model for composite $J=0$ mesons.

\section{First case: $\mathbf{SU(2)_R}$ is the replica of $\mathbf{SU(2)_L}$}

The $SU(2)_R$ generators  are given by (\ref{eq:G1}).

The eight $J=0$ composite representations of $SU(2)_R$ (which can be
decomposed into two doublets $(2)$ and $(\ol 2)$) are
built like for $SU(2)_L$; only the value of the mixing angle is different
$\varphi \not= \theta_c$:

\vbox{
\bea & &\Phi_R({\mathbb D},\varphi)= [{\mathbb M}\,^0, {\mathbb M}^3,
   {\mathbb M}^+, {\mathbb M}^-]_R({\mathbb D},\varphi)\cr & &\ \cr
&=& {\mathbb R}^\dagger(\varphi)
\left[  \frac{1}{\sqrt{2}}\left(\begin{array}{ccc}
      {\mathbb D} & \vline & 0\\
          \hline  0 & \vline & {\mathbb D}
                   \end{array}\right),
\frac{i}{\sqrt{2}} \left(\begin{array}{ccc}  {\mathbb D} & \vline & 0\\ 
\hline    0 & \vline & -{\mathbb D}
\end{array}\right),
i\left(\begin{array}{ccc} 0 & \vline & {\mathbb D}\\
    \hline   
 0 & \vline & 0           \end{array}\right),
 i\left(\begin{array}{ccc}  0 & \vline & 0\\ 
 \hline   {\mathbb D} & \vline & 0  \end{array}\right)
  \right]
{\mathbb R}(\varphi)\cr
&&\cr
& & =\left[  \frac{1}{\sqrt{2}}\left(\begin{array}{ccc}
      {\mathbb D} & \vline & 0\\
          \hline  0 & \vline & {{\cal R}}^\dagger(\varphi)\,{\mathbb
D}\,{{\cal R}}(\varphi)
                   \end{array}\right),
\frac{i}{\sqrt{2}} \left(\begin{array}{ccc}  {\mathbb D} & \vline & 0\\ 
\hline    0 & \vline & -{{\cal R}}^\dagger(\varphi)\,{\mathbb D}\,{{\cal
R}}(\varphi)
\end{array}\right),
\right. \cr
&& \left. \hskip 6cm
i\left(\begin{array}{ccc} 0 & \vline & {\mathbb D}\,{{\cal R}}(\varphi)\\
    \hline   
 0 & \vline & 0           \end{array}\right),
 i\left(\begin{array}{ccc}  0 & \vline & 0\\ 
 \hline    {{\cal R}}^\dagger(\varphi)\,{\mathbb D} & \vline & 0  \end{array}\right)
  \right].\cr
&&
\label{eq:repsR}
\eea
}

Their laws of transformations are given in \cite{Machet1}.
We use the same notation $\mathbb S$ for the scalar
mesons, $\mathbb P$ for the pseudoscalars, and make explicit the matrix
${\mathbb D} \in \{{\mathbb D}_1,{\mathbb D}_2,{\mathbb D}_3,{\mathbb
D}_4\}$ attached to the corresponding $({\mathbb S}^0_R,\vec{\mathbb P}_R)$ or
$({\mathbb P}_R^0,\vec{\mathbb S}_R)$ quadruplets, together with
the mixing angle $\varphi$.
For example ${\mathbb P}^+_R({\mathbb D}_3,\varphi)$ is the charged
pseudoscalar $J=0$ meson
belonging to the $({\mathbb S}^0_R,\vec{\mathbb P}_R)$ quadruplet of $SU(2)_R$
labelled by ${\mathbb D}_3$ and which depends on the mixing angle $\varphi$:
\begin{equation}
{\mathbb P}^+_R({\mathbb D}_3,\varphi) = i \left( \begin{array}{rrrr}
              & \vline & -\sin\varphi & \cos\varphi  \cr
              & \vline & \cos\varphi  & \sin\varphi \cr
                   \hline
              & \vline &  & \end{array} \right)_{pseudoscalar};
\label{eq:exR}
\end{equation}
the corresponding expression for $SU(2)_L$ is
\begin{equation}
{\mathbb P}^+_L({\mathbb D}_3,\theta_c) = i \left( \begin{array}{rrrr}
              & \vline & -\sin\theta_c & \cos\theta_c  \cr
              & \vline & \cos\theta_c  & \sin\theta_c \cr
                   \hline
              & \vline &  & \end{array} \right)_{pseudoscalar}.
\label{eq:exL}
\end{equation}
In general, unlike the Higgs multiplets introduced in other approaches
\cite{MohapatraSenjanovic}, the reps. of $SU(2)_R$ are not reps. of $SU(2)_L$
(and
vice-versa), the only exception occurring when $\varphi = \theta_c$.

\subsection{The gauge Lagrangian}

The kinetic terms for the mesons are built like in \cite{Machet1} by using
the property of
\footnote{Since throughout the paper we use a matricial {\em notation} for the
fields, the symbol $\otimes$ has been used to mean the product of fields
as space-time functions to avoid a possible misinterpretation with product
of matrices.}

\vbox{
\bea
{\cal J} &=&
({\mathbb S}^0, \vec {\mathbb P})({\mathbb D}_1)
        \otimes ({\mathbb S}^0, \vec {\mathbb P})({\mathbb D}_1) +
({\mathbb S}^0, \vec {\mathbb P})({\mathbb D}_2)
         \otimes ({\mathbb S}^0, \vec {\mathbb P})({\mathbb D}_2)\cr
 &+& ({\mathbb S}^0, \vec {\mathbb P})({\mathbb D}_3)
          \otimes ({\mathbb S}^0, \vec {\mathbb P})({\mathbb D}_3)
 - ({\mathbb S}^0, \vec {\mathbb P})({\mathbb D}_4)
           \otimes ({\mathbb S}^0, \vec {\mathbb P})({\mathbb D}_4) \cr
&-& ({\mathbb P}^0, \vec {\mathbb S})({\mathbb D}_1)
         \otimes ({\mathbb P}^0, \vec {\mathbb S})({\mathbb D}_1)
  - ({\mathbb P}^0, \vec {\mathbb S})({\mathbb D}_2)
          \otimes ({\mathbb P}^0, \vec {\mathbb S})({\mathbb D}_2)\cr
 &-& ({\mathbb P}^0, \vec {\mathbb S})({\mathbb D}_3)
         \otimes ({\mathbb P}^0, \vec {\mathbb S})({\mathbb D}_3)
 + ({\mathbb P}^0, \vec {\mathbb S})({\mathbb D}_4)
         \otimes ({\mathbb P}^0, \vec {\mathbb S})({\mathbb D}_4)
\label{eq:J}
\eea
}

to be invariant both \l
- by $SU(2)_L$ when the quadruplets $({\mathbb S}^0, \vec {\mathbb P})$,
  $({\mathbb P}^0, \vec {\mathbb S})$ are chosen to be the representations
of $SU(2)_L$, thus  expressed in terms of $\theta_c$;\l
- by $SU(2)_R$ when the quadruplets $({\mathbb S}^0, \vec {\mathbb P})$,
  $({\mathbb P}^0, \vec {\mathbb S})$ are chosen to be the representations
of $SU(2)_R$, thus expressed in terms of $\varphi$.

The two expressions for $\cal J$ can indeed be seen to be identical
 in the basis of quark "flavour eigenstates''
$\bar q_i q_j$ and $\bar q_i \gamma_5 q_j$ in which all dependence on the
mixing angle vanishes \cite{Machet1}.

Consequently, the kinetic terms for the mesons are obtained from $\cal J$ by
replacing each mesonic field by the corresponding covariant derivative with
respect to the $SU(2)_L \times SU(2)_R$ gauge group, and they can equally
be expressed using $SU(2)_L$ or $SU(2)_R$ quadruplets.

One introduces accordingly two sets of gauge bosons $\vec W_{\mu L}$ and
$\vec W_{\mu R}$,
and call the two corresponding coupling constants $g_L$ and $g_R$.

\subsection{The neutral gauge bosons}
\label{subsec:neutral1}

Neutral gauge bosons eventually get masses through kinetic terms
$\frac{1}{2}\, D_\mu {\mathbb P}^3 D^\mu {\mathbb P}^3$, where $D_\mu$ is
hereafter the covariant derivative with respect to $SU(2)_L \times SU(2)_R$;
indeed,
the axial part of the neutral gauge group generators, when acting on any 
neutral pseudoscalar ${\mathbb P}^3$, transforms it into a neutral scalar
\footnote{
The action of a right generator on a pseudoscalar described by the
matrix $\mathbb M$ ($[,]$ stands for the commutator and $\{,\}$ for the
anticommutator) is
\begin{equation}
{\mathbb T}^i_R . {\mathbb M}_{pseudoscalar} =
\frac{1}{2} \left( [{\mathbb M},{\mathbb T}_i]_{pseudoscalar}
             + \{{\mathbb M},{\mathbb T}^i\}_{scalar} \right)
\label{eq:Raction}
\end{equation}
and the action of a left generator
\begin{equation}
{\mathbb T}^i_L . {\mathbb M}_{pseudoscalar} =
\frac{1}{2} \left( [{\mathbb M},{\mathbb T}_i]_{pseudoscalar}
             - \{{\mathbb M},{\mathbb T}^i\}_{scalar} \right)
\label{eq:Laction}
\end{equation}
}
;
if the latter gets a non-vanishing vacuum expectation value (see subsection
\ref{subsec:break1} below), a mass term
arises for the corresponding neutral gauge field.

Though all four $({\mathbb S}^0,\vec{\mathbb P})({\mathbb D}_{1\cdots 4},
\theta_c)$ quadruplets are isomorphic to the complex scalar doublet of the
standard model, it is specially convenient, to ease the computations, to
choose the Higgs boson $H$ to be ${\mathbb S}^0({\mathbb D}_1)$ or
${\mathbb S}^0({\mathbb D}_4)$, which do not depend on the mixing angles;
this is also the case for the corresponding ${\mathbb P}^3$'s.

If one furthermore chooses $H$ to be $CP$-even, this restricts the Higgs
boson to $H = {\mathbb S}^0({\mathbb D}_1)$.

From
\bea
D_\mu {\mathbb P}_3({\mathbb D}_1) &\ni& \left(\p_\mu 
- ig_L W_{\mu L}^3 {\mathbb T}^3_L. - ig_R W_{\mu R}^3 {\mathbb
  T}^3_R.\right) {\mathbb P}_3({\mathbb D}_1) + \ldots\cr
&=& \p_\mu  {\mathbb P}_3({\mathbb D}_1) -
     \frac{1}{2}(g_L W_{\mu L}^3 - g_R W_{\mu R}^3) H + \ldots
\eea
and
\begin{equation}
< H > = < {\mathbb S}^0 ({\mathbb D}_1)> = \frac{v}{\sqrt{2}}
\end{equation}
one straightforwardly gets the spectrum of the neutral gauge fields:\l
- a massless photon 
\begin{equation}
A_\mu = \sin\theta_W W^3_{\mu L} +\cos\theta_W W^3_{\mu R}
\label{eq:Amu}
\end{equation}
- a massive $Z_\mu$
\begin{equation}
Z_\mu = \cos\theta_W W^3_{\mu L} - \sin\theta_W W^3_{\mu R}
\label{eq:Zmu}
\end{equation}
with mass
\begin{equation}
M_Z^2 = \frac{g^2 v^2}{16 \cos^2\theta_W},
\label{eq:MZ}
\end{equation}
and the usual relations
\begin{equation}
g_L = \frac{e}{\sin\theta_W},\ g_R = \frac{e}{\cos\theta_W},\ 
e = \frac{g_L g_R}{\sqrt{g_L^2 + g_R^2}}
\label{eq:gLgR}
\end{equation}
where $e$ is the unit electric charge.

In addition to the standard $Z_\mu-W^+_{\nu L}-W^-_{\rho L}$ coupling
proportional to $e \cos\theta_W/\sin\theta_W$, there exists now a coupling
$Z_\mu-W^+_{\nu R}-W^-_{\rho R}$ proportional to $e
\sin\theta_W/\cos\theta_W$, that is smaller than the previous one by a
factor $\tan^2\theta_W \approx .28$ (see also subsection \ref{subsec:pair}
below).

\subsection{The charged sector}

One selects in the kinetic terms the terms which give masses to the charged
$W$'s when $< H > \not= 0$. For $SU(2)_L$ alone,
the quadruplet $({\mathbb S}^0, \vec{\mathbb P})({\mathbb D}_1,\theta_c)$,
which includes the Higgs boson, is the only one concerned since
\bea
{\mathbb T}^+_L.{\mathbb P}^-({\mathbb D}_1,\theta_c) &\ni& -\frac{i}{\sqrt{2}}
{\mathbb S}^0({\mathbb D}_1) + \ldots \cr
{\mathbb T}^-_L.{\mathbb P}^+({\mathbb D}_1,\theta_c) &\ni& -\frac{i}{\sqrt{2}}
{\mathbb S}^0({\mathbb D}_1) + \ldots .
\eea
However, this quadruplet is not stable by the action of $SU(2)_R$
for $\varphi \not= \theta_c$, and gets mixed with
$({\mathbb S}^0, \vec{\mathbb P})({\mathbb D}_4,\theta_c)$.
More explicitly, one gets (keeping the terms depending on
${\mathbb S}^0({\mathbb D}_4)$ for the eventuality when it gets a
non-vanishing VEV too --see subsection \ref{subsec:break1} below--)
\bea
{\mathbb T}^+_R.{\mathbb P}^-({\mathbb D}_1,\theta_c) &\ni& \frac{i}{\sqrt{2}}
   \left(\cos(\varphi -\theta_c) {\mathbb S}^0({\mathbb D}_1)
+ \sin(\varphi-\theta_c){\mathbb S}^0({\mathbb D}_4)\right) + \ldots \cr
{\mathbb T}^-_R.{\mathbb P}^+({\mathbb D}_1,\theta_c) &\ni& \frac{i}{\sqrt{2}}
 \left(\cos(\varphi -\theta_c) {\mathbb S}^0({\mathbb D}_1)
- \sin(\varphi-\theta_c){\mathbb S}^0({\mathbb D}_4)\right) + \ldots \cr
{\mathbb T}^+_R.{\mathbb P}^-({\mathbb D}_4,\theta_c) &\ni& \frac{i}{\sqrt{2}}
   \left(- \sin(\varphi -\theta_c) {\mathbb S}^0({\mathbb D}_1)
+ \cos(\varphi-\theta_c){\mathbb S}^0({\mathbb D}_4)\right) + \ldots \cr
{\mathbb T}^-_R.{\mathbb P}^+({\mathbb D}_4,\theta_c) &\ni& \frac{i}{\sqrt{2}}
   \left( \sin(\varphi -\theta_c) {\mathbb S}^0({\mathbb D}_1)
+ \cos(\varphi-\theta_c){\mathbb S}^0({\mathbb D}_4)\right) + \ldots
\eea
One finds in
\hbox{$ D_\mu{\mathbb P}^+({\mathbb D}_1,\theta_c)\otimes
         D^\mu{\mathbb P}^-({\mathbb D}_1,\theta_c)      
- D_\mu{\mathbb P}^+({\mathbb D}_4,\theta_c) \otimes
   D^\mu{\mathbb P}^-({\mathbb D}_4,\theta_c)$}
the mass terms for the charged gauge bosons
\begin{equation}
\frac{v^2}{8}
\left(\begin{array}{cc} W_{\mu L}^- & W_{\mu R}^- \end{array}\right)
\left(\begin{array}{cc}
           g_L^2 & -g_L g_R\cos(\varphi - \theta_c) \cr
          -g_L g_R\cos(\varphi - \theta_c) & g_R^2  \end{array}\right)
\left(\begin{array}{c} W_{\mu L}^+ \cr W_{\mu R}^+ \end{array}\right)
\end{equation}
which corresponds to the two eigenvalues
\begin{equation}
M_W^2 = \frac{1}{2}
M_Z^2 \left( 1 \pm \sqrt{1-4 \sin^2\theta_W \cos^2\theta_W \sin^2(\varphi -
\theta_c)}\right).
\end{equation}
Owing to the constraint  that one of the eigenvalues has to match the
masses of the observed $W^\pm
_{\mu L}$'s
\begin{equation}
M^2_{W^\pm_{\mu L}} = g_L^2 \frac{v^2}{16} = \cos^2\theta_W M_Z^2
\end{equation}
one gets
\begin{equation}
\varphi = \theta_c + \frac{\pi}{2},
\label{eq:varphi}
\end{equation}
in which case the $W_L-W_R$ mixing vanishes and the mass eigenstates are the
electroweak eigenstates $W^\pm_{\mu L}$ and $W^\pm_{\mu R}$, with
\begin{equation}
M^2_{W^\pm_{\mu R}} = g_R^2 \frac{v^2}{16} = \sin^2\theta_W M^2_Z
\approx (43\,GeV)^2.
\label{eq:MWR}
\end{equation}
The case when one allows $< {\mathbb S}^0({\mathbb D}_4) > \not=0$
in addition to $< {\mathbb S}^0({\mathbb D}_1) > \not=0$  is a
straightforward generalization which leads
to the same spectrum for the gauge bosons: the only modification is that,
now, $ v^2/2 = < {\mathbb S}^0({\mathbb D}_1) > ^2 - <
{\mathbb S}^0 ({\mathbb D}_4)>^2$.

An immediate consequence of (\ref{eq:gLgR}) and (\ref{eq:MWR}) is that, while
at high energies  (in the $W_\mu$ propagator, $M_{W}^2$ can be neglected
with respect to $q^2$), the right-handed electroweak interactions are weaker
that the left-handed ones because $g_R < g_L$, their effective strengths
are identical  in the infrared regime $q^2 \ll M_W^2$ 
\begin{equation}
(\frac{g_L}{8M_{W_L}})^2 = (\frac{g_R}{8M_{W_R}})^2 =
\frac{e^2}{8\sin^2\theta_W \cos^2\theta_W M_Z^2} = \frac{G_F}{\sqrt{2}},
\label{eq:loweff}
\end{equation}
showing, in the charged sector, left-right symmetry  as a low energy phenomenon.
%
\subsection{Composite representations of $\mathbf{SU(2)_R}$}

The condition (\ref{eq:varphi}) enables the $J=0$ composite representations
(\ref{eq:repsR}) of $SU(2)_R$  to be rewritten as the quadruplets
\bea
\phi_{1R},\ti\phi_{1R} &=& \left[{\mathbb M}^0({\mathbb D}_1),
        {\mathbb M}^3({\mathbb D}_1),
        {\mathbb M}^+({\mathbb D}_4,\theta_c),
        {\mathbb M}^-({\mathbb D}_4,\theta_c)\right], \cr
\phi_{2R},\ti\phi_{2R} &=& \left[{\mathbb M}^0({\mathbb D}_4),
        {\mathbb M}^3({\mathbb D}_4),
        {\mathbb M}^+({\mathbb D}_1,\theta_c),
        {\mathbb M}^-({\mathbb D}_1,\theta_c)\right], \cr
\phi_{3R},\ti\phi_{3R} &=& \left[{\mathbb M}^0({\mathbb D}_2,\theta_c),
        {\mathbb M}^3({\mathbb D}_2,\theta_c),
        {\mathbb M}^+({\mathbb D}_3,\theta_c),
        {\mathbb M}^-({\mathbb D}_3,\theta_c)\right], \cr
\phi_{4R},\ti\phi_{4R} &=& \left[{\mathbb M}^0({\mathbb D}_3,\theta_c),
        {\mathbb M}^3({\mathbb D}_3,\theta_c),
        {\mathbb M}^+({\mathbb D}_2,\theta_c),
        {\mathbb M}^-({\mathbb D}_2,\theta_c)\right],
\label{eq:repsR1}
\eea
where, as usual, to a scalar ${\mathbb M}^0$ are associated three
pseudoscalar $\vec{\mathbb M}$'s, and vive-versa. We have emphasized above
that, ${\mathbb M}^{0,3}({\mathbb D}_1,{\mathbb D}_4)$ do not
depend on the mixing angle.

\subsection{Breaking $\mathbf{SU(2)_L \times SU(2)_R}$}
\label{subsec:break1}

We have only supposed up to now that two scalar fields, ${\mathbb
S}^0({\mathbb D}_1)$ and ${\mathbb S}^0({\mathbb D}_4)$  eventually get
non-vanishing vacuum expectation values, achieving the spontaneous breaking
of  $SU(2)_L \times SU(2)_R$ down to electromagnetic $U(1)_{em}$.

From (\ref{eq:repsR1}) the simplest natural potential invariant by
$SU(2)_R \times SU(2)_L$ which triggers
$< {\mathbb S}^0({\mathbb D}_1) >\not= 0$ and / or
$< {\mathbb S}^0({\mathbb D}_4) >\not= 0$, breaking it down to $U(1)_{em}$, is
\begin{equation}
V_1 = -\frac{\sigma ^2}{2}
\left(
({\mathbb S}^0,\vec {\mathbb P})^{\otimes 2}({\mathbb D}_1,\theta_c)
                  - ({\mathbb S}^0,\vec {\mathbb P})^{\otimes 2}({\mathbb
                    D}_4,\theta_c)
\right)
    + \frac{\lambda}{4}
\left(
({\mathbb S}^0,\vec {\mathbb P})^{\otimes 2}({\mathbb D}_1,\theta_c)
          - ({\mathbb S}^0,\vec {\mathbb P})^{\otimes 2}
        ({\mathbb D}_4,\theta_c)
\right) ^{\otimes 2},
\end{equation}
which has a non-trivial minimum for 
\begin{equation}
< {\mathbb S}^0({\mathbb D}_1)^{\otimes 2} - {\mathbb S}^0({\mathbb
D}_4)^{\otimes 2} >
    = < {\mathbb S}^0({\mathbb D}_1)^\dagger \otimes {\mathbb S}^0({\mathbb D}_1)
     + {\mathbb S}^0({\mathbb D}_4)^\dagger  \otimes{\mathbb S}^0({\mathbb
D}_4)>
      \not= 0.
\end{equation}
Since $< {\mathbb S}^0({\mathbb D}_4)> \not=0$ allows for an eventual
spontaneous violation of $CP$, it is natural to consider
$\vert< {\mathbb S}^0({\mathbb D}_4)>\vert \ll
\vert< {\mathbb S}^0({\mathbb D}_1)>\vert$.
One also imposes that no pseudoscalar can condensate in
the vacuum $<\vec {\mathbb P} ({\mathbb D}_{1\cdots 4} > =0$
(this condition can eventually be relaxed since pseudoscalar condensates
can a priori be generated by parity violating electroweak corrections).

Then, the six $\vec {\mathbb P}({\mathbb D}_1),\vec {\mathbb P}({\mathbb
D}_4)$ are classically massless.
For $< {\mathbb S}^0({\mathbb D}_4)>\approx 0$ the Higgs boson becomes
$h_1 \approx {\mathbb S}^0({\mathbb D}_1)-< {\mathbb S}^0({\mathbb D}_1)>$
with mass $M_H^2 = \lambda v^2$, while
$h_4 = {\mathbb S}^0({\mathbb D}_4)-< {\mathbb S}^0({\mathbb D}_4)>$
is classically massless.
Since there are only five massive gauge bosons $Z_\mu, W_{\mu L}^\pm,W_{\mu
R}^\pm$, ${\mathbb P}^3({\mathbb D}_4)$ and $h_4$ stay classically
massless. They however couple to a pair $W_{\mu L} W_{\mu R}$
with a coupling proportional to $g_L g_R < {\mathbb S}^0({\mathbb
D}_1)>$
and can acquire a small mass at the quantum level 
\footnote{It is not our subject here but the structure of composite
representations of $SU(2)_L$ and $SU(2)_R$ allows to write simple invariant
mass terms for the other mesons not involved in the symmetry breaking
potential \cite{Machet1}.}.

More general symmetry breaking potentials involving other quadruplets can
be used, which in
general increases the number of pseudo-goldstone bosons (see next section).

\section{Second case: the ``inverted'' $\mathbf{SU(2)_R}$}

The first chiral extension of the Standard Model studied above is very rigid:
because the mixing angle of the right sector is fixed by (\ref{eq:varphi}),
all couplings of gauge fields to $J=0$ mesons are fixed too
, with reduced hope to fit to experimental constraints.
And, indeed, as will be shown in the last section, the first extension above
does not seem to be a suitable one.

This is why we now investigate the second possibility
$SU(2)_R = {\cal G}_{2R}$ with generators given by (\ref{eq:G2}) in which,
in particular, no constraint arises for the mixing angle in the right sector,
allowing the tuning of the $W^\pm_{\mu R}$ gauge bosons couplings.

The equivalent of the Cabibbo matrix is ${\mathbb D}_2 {\mathbb C}$ (with
determinant $-1$) and the $J=0$ (sum of $(2)$ and $(\ol 2)$ doublets) 
composite representations  of ${\cal G}_{2R}$ are the quadruplets
\bea
\chi_{1R},\tilde\chi_{1R} &=& \left[{\mathbb M}^0({\mathbb D}_1),
        {\mathbb M}^3({\mathbb D}_1),
        {\mathbb M}^+({\mathbb D}_2,\varphi),
        {\mathbb M}^-({\mathbb D}_2,\varphi)\right], \cr
\chi_{2R},\tilde\chi_{2R} &=& \left[{\mathbb M}^0({\mathbb D}_2,\varphi),
        {\mathbb M}^3({\mathbb D}_2,\varphi),
        {\mathbb M}^+({\mathbb D}_1,\varphi),
        {\mathbb M}^-({\mathbb D}_1,\varphi)\right], \cr
\chi_{3R},\tilde\chi_{3R} &=&\left[{\mathbb M}^0({\mathbb D}_3,\varphi),
        {\mathbb M}^3({\mathbb D}_3,\varphi),
        {\mathbb M}^+({\mathbb D}_4,\varphi),
        {\mathbb M}^-({\mathbb D}_4,\varphi)\right], \cr
\chi_{4R},\tilde\chi_{4R} &=& \left[{\mathbb M}^0({\mathbb D}_4),
        {\mathbb M}^3({\mathbb D}_4),
        {\mathbb M}^+({\mathbb D}_3,\varphi),
        {\mathbb M}^-({\mathbb D}_3,\varphi)\right],
\label{eq:repsR2}
\eea
all of them being either of the type $ \chi =({\mathbb S}^0, \vec{\mathbb P})$
or $\tilde\chi =({\mathbb P}^0,\vec{\mathbb S})$. We have again emphasized in
(\ref{eq:repsR2}) that ${\mathbb M}^{0,3}({\mathbb D}_1,{\mathbb D}_4)$ in
fact do not depend on $\varphi$.

One takes for the kinetic terms the same expression deduced from the invariant
${\cal J}$ (\ref{eq:J}) as in the first extension, since ${\cal J}$, which
is diagonal in the basis of flavour eigenstates (all dependence on the
mixing angles disappears then), can equally be  expressed in terms of the
$\chi$'s and $\ti\chi$'s.

\subsection{The neutral gauge bosons}
\label{subsec:neutral2}

Nothing is changed with respect to the discussion made in the previous
case, because ${\mathbb T}^3_2 \equiv {\mathbb T}^3_1$, and  one gets again
eqs.~(\ref{eq:Amu}), (\ref{eq:Zmu}), (\ref{eq:MZ}) and (\ref{eq:gLgR}).
The $Z_\mu-W^+_{\nu R}-W^-_{\rho R}$ coupling is also the same as
previously.

\subsection{The charged sector}

Choosing again the Higgs field to be $H = {\mathbb S}^0({\mathbb D}_1)$,
it is simple matter to realize from (\ref{eq:repsR2}) that the mass terms for
the charged gauge fields are again diagonal (no $W_L-W_R$ mixing);
they are generated by acting on charged mesons with left or right generators
such that the result of this action is ${\mathbb S}^0({\mathbb D}_1)$;
the $W_{\mu L}^\pm$'s get their
masses when acting with ${\mathbb T}^\pm_L$ on ${\mathbb P}^\mp({\mathbb
D}_1)$ and the $W_{\mu R}^\pm$'s when acting with ${\mathbb T}^\pm_R$ on
${\mathbb P}^\mp({\mathbb D}_2)$.
Thus, if one writes the
Lagrangian in the basis of the $\chi, \ti\chi$'s, $(1/2)D_\mu \chi_1 D^\mu
\chi_1^\dagger$ gives masses to $W_{\mu R}^\pm$ and
$(1/2)D_\mu \chi_2 D^\mu \chi_2^\dagger$ to $W_{\mu L}^\pm$,
whatever be the value of the mixing angle $\varphi$.

One finds, like in the replica case, that
\begin{equation}
M_{W_{\mu L}^\pm} = \cos\theta_W M_Z,\quad M_{W_{\mu R}^\pm} = \sin\theta_W
M_Z;
\end{equation}
the relation (\ref{eq:loweff}) implementing left-right symmetry at low
energy is still verified.

The difference is that, now, the mixing angle
$\varphi$ is left arbitrary and, in particular, has no connection with
the Cabibbo angle.

\subsection{Breaking $\mathbf{SU(2)_L \times SU(2)_R}$ down to
$\mathbf{U(1)_{em}}$}

From (\ref{eq:repsR2}), a natural $SU(2)_R$ invariant potential to trigger
$< {\mathbb S^0}({\mathbb D}_1)> \not= 0$ is
\bea
V_{2R}
       &=& \chi_{1R}^{\otimes 2} + \chi_{2R}^{\otimes 2}\cr
       &=& - \frac{\sigma^2}{2}\left(
           ({\mathbb S}^0,\vec {\mathbb P})^{\otimes 2}({\mathbb D}_1,\varphi)
 + ({\mathbb S}^0,\vec {\mathbb P})^{\otimes 2}({\mathbb D}_2,\varphi)\right)
 + \frac{\lambda}{4}\left(
({\mathbb S}^0,\vec {\mathbb P})^{\otimes 2}({\mathbb D}_1,\varphi)
+ ({\mathbb S}^0,\vec {\mathbb P})^{\otimes 2}({\mathbb D}_2,\varphi)\right)
^{\otimes 2}\cr
&&
\eea
but, for $\varphi \not= \theta_c$, it is not invariant by $SU(2)_L$;
in this general case, it appears that
$\cal J$ (\ref{eq:J}) is the only expression quadratic in the fields which is
left-right invariant; one is accordingly led to introduce the scalar potential
\begin{equation}
{\cal V} = -\frac{\sigma ^2}{2} {\cal J} + \frac{\lambda}{4} {\cal
J}^{\otimes 2}.
\end{equation}
${\cal J}$, and thus $\cal V$ too, is invariant by the full chiral group
$U(N)_L \times U(N)_R$
\footnote{To demonstrate this, it is convenient
to work in the basis of ${\cal P}_{even}$ and ${\cal P}_{odd}$
\cite{Machet1}
flavour eigenstates, and operate on them with the $U(N)$ generators
expressed in this same basis (having only one non-vanishing entry equal to
$1$).}
.

If one imposes
\begin{equation}
< {\mathbb P}^{0, 1, 2, 3}({\mathbb D}_{1\cdots 4})> = 0
\end{equation}
$\cal V$ has a non-trivial minimum for
\bea
&&< {\mathbb S}^0({\mathbb D}_1)^{\otimes 2} + {\mathbb S}^0({\mathbb
D}_2)^{\otimes 2} +
{\mathbb S}^0({\mathbb D}_3)^{\otimes 2} - {\mathbb S}^0({\mathbb
D}_4)^{\otimes 2}\cr
&& \hskip 3cm -{\mathbb S}^3({\mathbb D}_1)^{\otimes 2} - {\mathbb S}^3({\mathbb
D}_2)^{\otimes 2} -
{\mathbb S}^3({\mathbb D}_3)^{\otimes 2} + {\mathbb S}^3({\mathbb
D}_4)^{\otimes 2} >
=\frac{v^2}{2} \not= 0.
\label{eq:vac}
\eea
The above combination of neutral scalars  does not depend
on the mixing angle and can be evaluated with quadruplets of $SU(2)_L$ or
$SU(2)_R$.

We are led not to limit ourselves to the case when only $<{\mathbb
S}^0({\mathbb D}_1)> \not= 0$ but to consider the general case
\bea
&&< {\mathbb S}^0({\mathbb D}_1)> \not=0, < {\mathbb S}^0({\mathbb
D}_2)> \not=0, < {\mathbb S}^0({\mathbb D}_3)> \not=0,
< {\mathbb S}^0({\mathbb D}_4)> \not=0,\cr
&&< {\mathbb S}^3({\mathbb D}_1)> \not=0, < {\mathbb S}^3({\mathbb
D}_2)> \not=0, < {\mathbb S}^3({\mathbb D}_3)> \not=0,
< {\mathbb S}^3({\mathbb D}_4)> \not= 0,
\label{eq:vacuum}
\eea
with the ${\mathbb S}^0$'s and ${\mathbb S}^3$'s belonging indifferently to
left (depending on $\theta_c$) or right (depending on $\varphi$) quadruplets.
(\ref{eq:vacuum}) eventually switches on spontaneous $CP$ violation
\footnote{It is trivial to
restore all signs to be positive in (\ref{eq:vac}), (\ref{eq:vacuum})
by introducing (or removing) suitable ``$i$'' factors.}
.
Accidental additional invariance of the vacuum in the broken phase may arise.

At the minimum of the potential, the mass terms for all
pseudoscalars and charged scalars vanish (we suppose that
$< {\mathbb M}_i {\mathbb M}_j > =<
{\mathbb M}_i> < {\mathbb M}_j>,\ \forall {\mathbb M}_{i,j}$).
The spectrum of neutral scalars is obtained by diagonalizing their mass matrix
${\cal M}^2$, the entries of which are the corresponding products of  vacuum
expectations values, for example
\begin{equation}
{\cal M}^2_{{\mathbb S}^0({\mathbb D}_1){\mathbb S}^0({\mathbb D}_2)} =
\lambda < {\mathbb S}^0({\mathbb D}_1)>< {\mathbb S}^0({\mathbb D}_2)>.
\end{equation}
The eigenvalues of ${\cal M}^2$, independent of the mixing angles
$\theta_c$ and $\varphi$, are all vanishing but one equal to
$M_h^2 = \lambda v^2$ .
In the special case when only $<{\mathbb S}^0({\mathbb D}_1)> \not= 0$,
the Higgs boson is $h = h_1 \equiv {\mathbb S}^0({\mathbb D}_1) - <{\mathbb
S}^0({\mathbb D}_1)>$.

The mesonic spectrum which then arises is the following:\l
- 1 neutral and 4 charged pseudoscalar goldstones which get absorbed by the 5
massive gauge bosons;\l
- $N^2 -5 = 11$ pseudoscalars pseudo-goldstone bosons;\l
- one Higgs boson $h$ with mass $M_h^2 = \lambda v^2$;\l
- $N^2 -1 = 15$ scalar pseudo-goldstone bosons.

(\ref{eq:vacuum}) corresponds to the spontaneous breaking of
$U(N)_L \times U(N)_R$ down to $U(1)_{em}$. Because the kinetic and gauge
terms in the Lagrangian do not have the full $U(N)_L \times U(N)_R$ but
only the $SU(2)_L \times SU(2)_R$ chiral invariance, among the
$2N^2 -1$ goldstone bosons expected in the breaking triggered by $\cal V$,
only five true goldstones, which correspond to the breaking of the gauge
subgroup $SU(2)_L \times SU(2)_R$, are eaten by the five massive gauge fields;
the other $2N^2 -6$ are pseudo-goldstones which are expected to get massive
only by quantum effects, which makes likely a hierarchy between the
electroweak / Higgs mass scales and theirs.

{\em Remark}: (\ref{eq:vacuum}) has consequences on leptonic decays of $J=0$
mesons mediated by $W^\pm_{\mu L}$ (we neglect the ones mediated by
$W^\pm_{\mu R}$, supposing the right-handed neutrinos to be heavy in
association with a see-saw mechanism involving a high mass scale);
however, up to small deviations which can be
accounted for by variations in the leptonic decay constants, the dependence of
the decay amplitudes on the Cabibbo angle is very well explained
\cite{Machet1} by
supposing that ${\mathbb S}^0({\mathbb D}_1)$ is the only scalar
condensing in the vacuum and that, accordingly, $h=h_1$. One is thus led
to consider that all other scalar condensates are much smaller than
$<{\mathbb S}^0({\mathbb D}_1)>$.

\section{Detection of $\mathbf{W^\pm_{{\bs \mu} R}}$}

We show here that the charged gauge fields of the right sector are likely
to have escaped detection.

\subsection{Hadronic colliders}

If $W^\pm_{\mu R}$ exist with a mass $\approx 43\,GeV$, they have to be produced
in hadronic interactions and specially at proton colliders.
However past experiments cannot have detected them; indeed, the decay
of $W^\pm_{\mu R}$ into two jets was only investigated by the UA2
experiment \cite{UA2} at CERN; unfortunately, their lower threshold was
$M_{2\,jets} > 48\,GeV$.

Let us also calculate the approximate  width of $W^\pm_{\mu R}$.

The picture that we have adopted here has given up the quarks as fundamental
fields; however, since the starting $SU(2)_L \times U(1)$ Lagrangian for
mesons is built with full compatibility with the standard model for quarks,
it is legitimate to consider that its chiral extension should also have its
counterpart at the quark level, at least as far as the couplings of the
charged gauge bosons are concerned
\footnote{The case of neutral current is more subtle since $U(1)_{\mathbb
I}$ gets involved as soon as fermions are concerned.}
. We accordingly consider that the two
$SU(2)_R$ groups ${\cal G}_{1R}$ and ${\cal G}_{2R}$ provide the two natural
chiral extensions of the Glashow-Salam-Weinberg model for quarks in the
charged current sector, with $\Psi$ (\ref{eq:psi}) in their fundamental
representation.

A good estimate of the width of $W^\pm_{\mu L}$ is obtained from the one-loop
self-energy from leptons and quarks, and, for the latter, considering only
the $u$ and $d$
quarks provides a reasonable approximation (which is all the more valid in
the right sector as the top quark is then above threshold):
\begin{equation}
\Gamma ^{total}_{W^\pm_L}
\approx (\sqrt{2}/\pi) G_F M^3_{W_L} \equiv \frac{g_L^2}{4\pi} M_{W_L}
\approx 2.5\,GeV.
\end{equation}
One gets an upper bound for the width of the $W^\pm_{\mu R}$ gauge bosons by
scaling the previous formula using the relations $M_{W_R}=
(\sin\theta_W/\cos\theta_W) M_{W_L}$
and $g_R =  (\sin\theta_W/\cos\theta_W)g_L$:
\begin{equation}
\Gamma^{total}_{W_R} \leq \frac{g_R^2}{4\pi} M_{W_R}
\equiv \left( \frac{\sin\theta_W}{\cos\theta_W}\right)^3
\Gamma^{total}_{W_L} \approx 200\,MeV;
\label{eq:Rwidth}
\end{equation}
indeed, the leptonic decays should be subtracted from this estimate, which
also supposes that the mixing angle to the first generation of quarks is
the Cabibbo angle; for ${\cal G}_1$, it is clearly an overestimate since
$\sin\theta_c$ is now involved  instead of $\cos\theta_c$ for the left gauge
bosons; for ${\cal G}_2$, this amounts to taking $\cos\varphi =
\cos\theta_c \approx .975$, very close to its absolute upper bound, which
is good enough for our purposes.

(\ref{eq:Rwidth}) is clearly a very small value for $\Gamma^{total}_{W_R}$
\begin{equation}
\frac{\Gamma^{total}_{W_R}}{M_{W_R}} \approx 4.6\,10^{-3},\quad \ll \quad
\frac{\Gamma^{total}_{W_L}}{M_{W_L}} \approx 2.5\,10^{-2},
\end{equation}
making all the more difficult an eventual detection.

Hadron colliders are thus certainly not a good place to look for the
$W^\pm_{\mu R}$'s.

\subsection{Pair production}
\label{subsec:pair}

As mentioned previously (see subsections \ref{subsec:neutral1},
\ref{subsec:neutral2}), the coupling of a pair of $W_R$'s to the massive
$Z_\mu$ is damped by a factor $\tan^2\theta_W$ with respect to its
equivalent for $W_L$'s; to this damping has to be added the one
occurring in leptonic decays of the $W_R$'s because of the presumably
heaviness of the right-handed neutrinos, advocating for a see-saw mechanism
\cite{BilenkyGiuntiGrimus} with another very high mass scale;
this concurs to make the detection
of a pair of $W_R$'s in $e^+e^-$ colliders very difficult and unlikely.

\subsection{Electroweak $\mathbf{SU(2)_R}$ interactions}

One must next investigate whether the right gauge bosons can be detected
through specific electroweak processes.

\subsubsection{Hadronic decays of pseudoscalar mesons}

Since the $W^\pm_{\mu R}$ are presumably not expected to be detectable
through their
leptonic decays because of the heavy masses of right-handed neutrinos,
 one should consider hadronic processes.

Now, the couplings of the right gauge bosons to flavour eigenstates (that we
consider, like in \cite{Machet1}, to be the asymptotic states) depend on a
mixing angle which can be, in the case of replica $SU(2)_R$,
$\varphi = \theta_c + \pi/2$ and, in the case of inverted $SU(2)_R$, an
arbitrary $\varphi$ to be determined experimentally like the Cabibbo angle for
left-handed interactions.

We shall be concerned with the decays of the type $P \rightarrow P_1P_2P_3$
where the $P$'s are pseudoscalar mesons.
In the quark picture the corresponding diagrams can be cast into two subsets:
the ``factorizable'' ones (fig.~1), which can be divided into two
disconnected parts by cutting the internal $W^\pm_{\mu}$ line, and the
``non-factorizable'' ones (fig.~2), which cannot.

\vbox{
\begin{center}
\epsfig{file=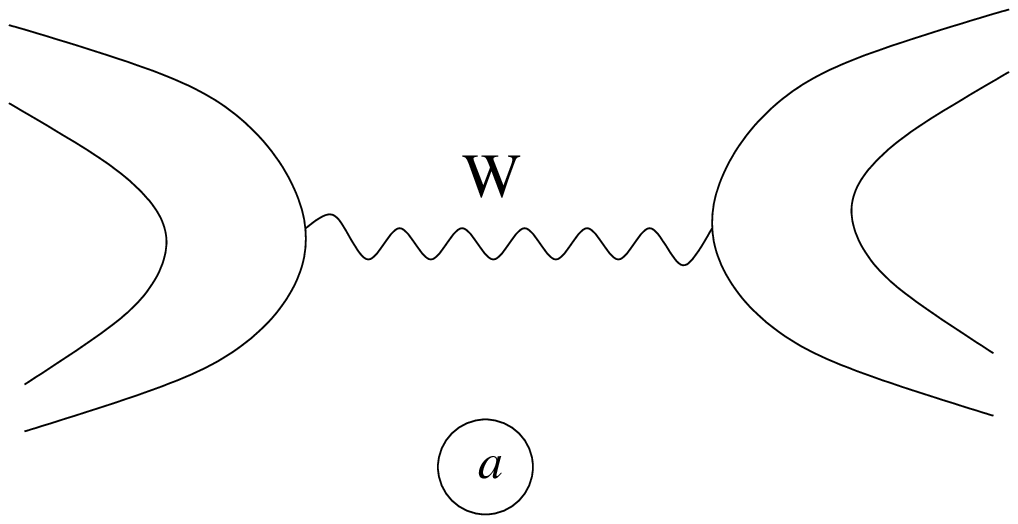,height=3truecm,width=6truecm}
\hskip 2cm
\epsfig{file=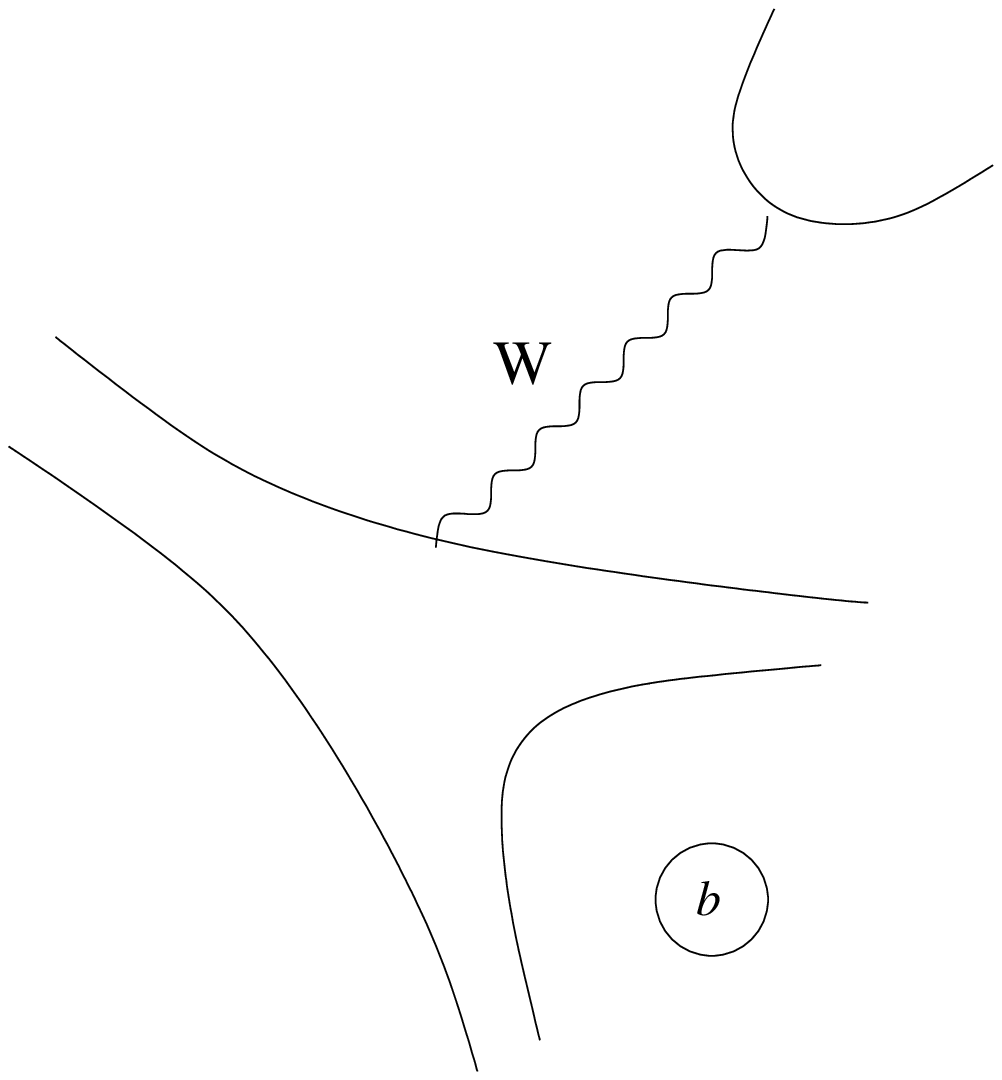,height=5truecm,width=6truecm}
\figskip
{\em Fig.~1: Factorizable contributions to $P \rightarrow P_1P_2P_3$ in the
quark picture.}
\end{center}
} 
\vbox{
\begin{center}
\epsfig{file=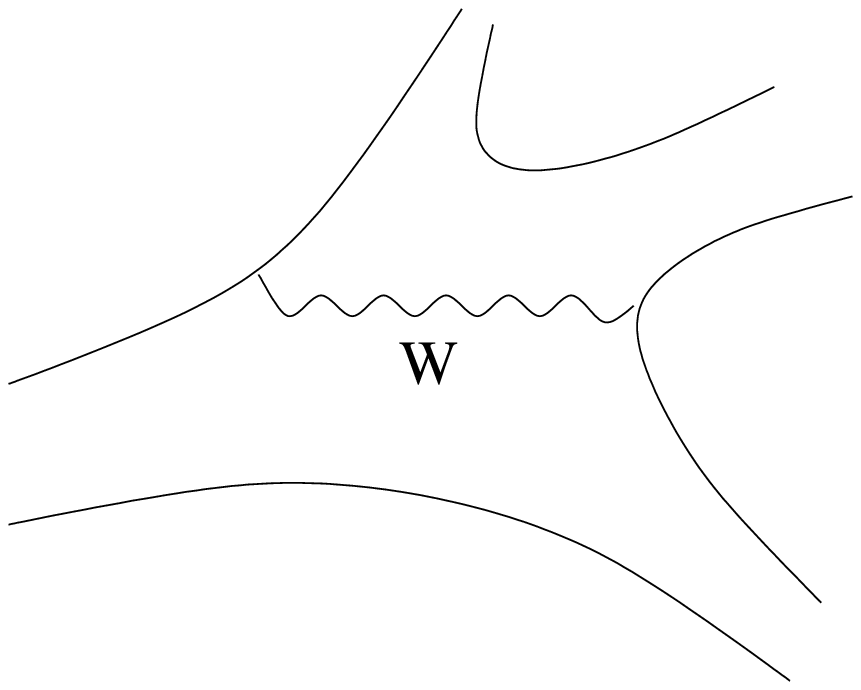,height=4truecm,width=6truecm}
\figskip
{\em Fig.~2: Non-factorizable contributions to $P \rightarrow P_1P_2P_3$ in the
quark picture.}
\end{center}
} 

Most decays involve both factorizable and non-factorizable contributions,
for which an eventual cancellation between $W_L$ and $W_R$ is 
difficult to evaluate; indeed, in the quark picture, one is led to
introducing intricate ``QCD'' corrections, and, in our approach,
a new type of ``strong-like'' interaction like in \cite{Machet3}, which, in the case of
a 3-body final states like the present one, enormously increases the number
of diagrams to take into account.

We are thus inclined to look for decays
which involve only factorizable contributions. This is the case for the decay
\begin{equation}
D^+ \rightarrow \pi^0 K^+ \pi^0
\end{equation}
which is described respectively at the quark level by the diagrams of
fig.~1 and in our model by fig.~3.

\vbox{
\begin{center}
\epsfig{file=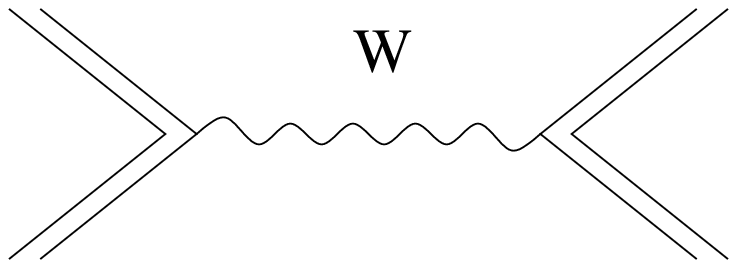,height=2truecm,width=5truecm}
\figskip
{\em Fig.~3: The decay $P \rightarrow P_1P_2P_3$ in the present model for
$J=0$ mesons.}
\end{center}
}

There are two types of factorizable diagrams: the first (fig.~1a) would
correspond to the formation of a $K^\ast$ resonance then decaying into
$K\pi$, and the second (fig.~1b) to that of a $\rho$ resonance which
yields two $\pi$'s; for the latter, the outgoing $K$ is akin to the
longitudinal part of the massive $W^\pm_\mu$ \cite{Machet1}.

It seems legitimate to make a one-to-one correspondence between the
factorizable diagrams of fig.~(1a) and the ones of fig.~(3) which naturally
arise, in the proposed extension of the standard model to $J=0$, from the
derivative couplings of one gauge fields to two pseudoscalars; the
transcription of the diagrams of figs.~(1b) and (2) needs introducing 
another type of interactions \cite{Machet3}.

While the $\rho^0 K^+$ channel is mentioned, the $\pi^0 K^+
\pi^0$ does not appear in the Table of Particle Properties
\footnote{There only appears the similar doubly Cabibbo suppressed decay
$D^+ \rar K^+ \pi^+ \pi^-$ which corresponds in the quark model to
non-factorizable contributions (see fig.~2).}
.

We interpret this as the absence of  the  channel associated with fig.~3,
and  attribute it to  a cancellation between the contributions of
$W^\pm_{\mu L}$ and $W^\pm_{\mu R}$; this distinguishes between
the two types of chiral extensions that we proposed.
Staying in the low energy limit where one can neglect the momentum
dependence of the $W^\pm_\mu$ propagator, it is trivial matter to deduce,
from the expression of the generators
\begin{equation}
{\mathbb T}^+_{L}(\varphi) = \left(\begin{array}{rrrr}
      & \vline &  \cos\theta_c & \sin\theta_c \cr
      & \vline & -\sin\theta_c & \cos\theta_c \cr
                    \hline
      & \vline & & \end{array}\right),
\end{equation}
${\mathbb T}^+_{1R}(\varphi=\theta_c + \pi/2)$ given by (\ref{eq:G1}) and
${\mathbb T}^+_{2R}(\varphi)$ given by (\ref{eq:G2})
that the
dependence of the amplitude on the mixing angle, which is $\sin^2\theta_c$
for $W^\pm_L$ alone becomes:\l
- $\sin^2\theta_c + \cos^2\theta_c = 1$ if one takes $SU(2)_R \equiv {\cal
  G}_1$;\l
- $\sin^2\theta_c - \sin^2\varphi$ if one takes $SU(2)_R \equiv {\cal G}_2$.

This decay would thus be strongly enhanced  if $SU(2)_R \equiv
{\cal G}_1$, since there is no more Cabibbo suppression; we are thus
led to favour the inverted $SU(2)_R$ as the preferred extension of our
model; small values of the mixing angle $\varphi$ are obviously favoured
to avoid a large contribution from the right sector; this statement will
get strengthened in the following subsection.

\subsubsection{The $\mathbf{K_L-K_S}$ mass difference}

The most stringent lower bounds for the mass of $W^\pm_{\mu R}$ come from
their contribution to the box diagrams controlling the $K_L-K_S$ mass
difference \cite{BeallBanderSoni}. They have however been obtained with
restrictive
hypothesis concerning in particular the coupling and mixing angle in the
right sector. Different conclusions can be reached in the framework
proposed above.

We consider here again that the extension that we proposed for $J=0$ mesons
can be transcribed at the level of quarks in the most intuitive way for
charged currents.

Including the contributions of $W^\pm_{\mu R}$, the box diagrams evaluated
with $u$ and $c$ quarks
\footnote{The role of the top quark is expected to be small as far as the
$K_L-K_S$ mass difference is concerned, because of the behaviour of the
corresponding mixing angles \cite{GavelaMachetPetcov}}
, in the 't Hooft-Feynman gauge
\footnote{They have been shown, in the quest for gauge invariance
\cite{HouSoniBasecqLiPal}, to yield the dominant contribution.}
 and neglecting
the momenta of external quarks for the gauge fields, yield the amplitude
\begin{equation}
{\cal A}(d\bar s \rar s \bar d) = {\cal A}_{LL} \left(
1 + \frac{\sin^2\varphi \cos^2\varphi}{\sin^2\theta_c \cos^2\theta_c} 
     \frac{{\cal O}_{RR}}{{\cal O}_{LL}}
+ \frac{\sin\varphi \cos\varphi}{\sin\theta_c \cos\theta_c}
      \frac{(M_{W_L}^2/M_{W_R}^2)\ln(M_{W_L}^2/M_{W_R}^2)}
               { 1-M_{W_L}^2/M_{W_R}^2}
     \frac{{\cal O}_{LR}}{{\cal O}_{LL}} \right),
\end{equation}
where ${\cal A}_{LL}$ corresponds to the standard result with two $W^\pm_{\mu
L}$'s
\begin{equation}
{\cal A}_{LL} = \frac{G_F^2}{4\pi^2} m_c^2 \sin^2\theta_c \cos^2\theta_c
                \ {\cal O}_{LL},
\end{equation}
the second term in the parenthesis corresponds to the contributions
of two $W^\pm_{\mu R}$'s, the third term to the $W^\pm_{\mu L}-W^\pm_{\mu R}$
crossed contributions,
and ${\cal O}_{LL}$, ${\cal O}_{RR}$ and ${\cal O}_{LR}$ are respectively
the operators
\bea
{\cal O}_{LL} &=& [\bar s \gamma_\mu(1-\gamma_5) d]\ 
                        [\bar s \gamma_\mu(1-\gamma_5) d],\cr
{\cal O}_{RR} &=& [\bar s \gamma_\mu(1+\gamma_5) d]\ 
                        [\bar s \gamma_\mu(1+\gamma_5) d],\cr
{\cal O}_{LR} &=& [\bar s (1-\gamma_5) d]\ 
                        [\bar s (1+\gamma_5) d],
\eea
the matrix elements of which between $\bar K^0$ and $K^0$ we approximate as
usual by inserting the vacuum as the intermediate state and using $PCAC$;
this leads to 
\begin{equation}
< \bar K^0 \vert {\cal O}_{RR} \vert K^0> =
      < \bar K^0 \vert {\cal O}_{LL} \vert K^0>    
\end{equation}
and to \cite{BeallBanderSoni}
\begin{equation}
< \bar K^0 \vert {\cal O}_{LR} \vert K^0> \approx
          7.7 < \bar K^0 \vert {\cal O}_{LL} \vert K^0>.
\end{equation}
One finally gets for $M_{W_{\mu L}} \approx 80\,GeV$ and
$M_{W_{\mu R}}\approx 43\,GeV$
\begin{equation}
{\cal A}(K^0 \rar \bar K^0) \approx {\cal A}_{LL}(K^0 \rar \bar K^0) \left(
 1 + \frac{\sin^2\varphi \cos^2\varphi}{\sin^2\theta_c \cos^2\theta_c}
   - 13.45 \frac{\sin\varphi \cos\varphi}{\sin\theta_c \cos\theta_c}\right)
\label{eq:KLKS}
\end{equation}
The corrections due to $W^\pm_{\mu R}$ vanish with $\varphi$, which is
constrained by (\ref{eq:KLKS}) to
\begin{equation}
\varphi \ll \theta_c,
\end{equation}
and the mixing angle in the right sector can be chosen small enough for
$W^\pm_{\mu R}$ to give negligeable contributions to the $K_L-K_S$ mass
difference.
 
\subsubsection{$\mathbf{{\bs \mu} \rightarrow e {\bs \gamma}}$}

The smallness of the
mixing angles in the right sector can counterbalance as much as desired the
small mass of the right gauge bosons and prevent any unwanted enhancement
of such decays \cite{MohapatraSenjanovic}.

\subsection{Light but elusive $\mathbf{W^\pm_{{\bs \mu} R}}$'s}

We have shown that a chiral completion of the standard electroweak model
in the mesonic sector can call for $W^\pm_{\mu R}$ gauge bosons much
lighter than usually expected  which, nevertheless, are likely to have
escaped detection.

In the extension favoured by experimental data  the two families of
right-handed quarks lie in inequivalent representations of $SU(2)_R$
and the associated mixing angle  is constrained to be much smaller that the
Cabibbo angle.

\section{Conclusion}
\label{section:conclusion}

The large arbitrariness in completing the standard model with a right-handed
sector was often reduced with arbitrary choices, including the
equality of the coupling constants \cite{SenjanovicMohapatra} and simple
relations between the left and right mixing angles \cite{LangackerSankar}.
Also, when dealing with fermions, the $U(1)_{\mathbb Y}$ of weak hypercharge
cannot be embedded in a right handed group, making an extra  massive $Z'_\mu$
 gauge field $Z'_\mu$ expected, in addition to new massive $W^\pm_R$'s. 
Last, the universality of behaviour for the different families of fermions
observed for left-handed interactions were always assumed to be also true
for right-handed interactions.

The three points mentioned above have received here  somewhat less
conventional answers:\l
- the existence of an extra $Z_\mu'$ has been decoupled from the spectrum
of the charged $W^\pm_{\mu R}$ by working in the (composite)
mesonic sector, which includes in particular the Higgs multiplet(s)
responsible of the gauge boson masses;\l
- the right coupling constant is different from the left one, as
  suggested by the chiral Gell-Mann-Nishijima relation and, in no case,
 the mixing angles in the right and left sectors appear to
  match; the former is most likely (still) another arbitrary parameter
to be determined from experiment, excluding  in particular the cases of a
manifest or pseudo-manifest left-right symmetry \cite{LangackerSankar}; \l
- the universality in the behaviour of families for left-handed interactions
  cannot be taken for granted in the right sector and is likely
to be broken in a very specific way if the independence of the left and right
mixing angles, yielding the maximum flexibility of the model,
is to be achieved (which seems indeed to be the situation favoured by
experimental data).

Problems arise when one wants to use this approach for leptons,
since the $U(1)_{\mathbb I}$ group  cannot be concealed any more.
A large variety of $SU(2)_L \times SU(2)_R \times U(1)_{\mathbb I}$
$(2,\ol2,1)$ multiplets, formed from doublets making up the reduceable
composite $SU(2)_L$ and $SU(2)_R$ quadruplets exhibited above, can be used to
give masses to the fermions (and then the hierarchy of fermionic masses has
to be linked to a hierarchy between different ``$q \bar q$ condensates''),
but a coupling of $W^3_{\mu R}$ to $ -(1/2)\bar\Psi\gamma^\mu{\mathbb I}\Psi$
which arises if we keep matching $B_\mu$ with $W^3_{\mu R}$ in the
Gell-Mann-Nishijima relation without further modification explicitly breaks
$SU(2)_R$. The perspective of adding an extra $Z'_\mu$
\footnote{Many studies have already dealt with the physics of $Z'_\mu$
gauge bosons; see for example \cite{Z'}
and references therein.}
appears non trivial: it cannot in particular be coupled to $U(1)_{\mathbb I}$
alone without extending the Higgs structure of the model, since  none
of the quark-antiquark bound states of the type (\ref{eq:repsL}) or
(\ref{eq:repsR}), being invariant by the action of $\mathbb I$, can make it
massive
\footnote{To give mass to a $Z'_\mu$ only coupled to $U(1)_{\mathbb I}$ one
needs for example to introduce new Higgs multiplets which are not
$U(1)_{\mathbb I}$ singlets, like $SU(2)$ triplets with nonvanishing $B-L$.
These can in particular trigger the wished  for see-saw mechanism
, but are usually associated with charged $W^\pm_{\mu R}$ much heavier
than considered here
.}
.
We thus have to tackle a whole reconstruction of the extended model,
probably also influencing the charged $W^\pm_{\mu R}$, suitable for the
leptonic sector, but  which has to accommodate
the outcome of the present study in the (composite) mesonic sector.
The challenge of generating reasonable neutrino masses (including having
a large scale to trigger a see-saw mechanism as invoked in this work)
and avoiding cosmological problems also comes into play.

This will be the subject of a subsequent work.
I hope that this limited study has suggested that considering interactions
between quark-antiquark composite fields used as reasonable substitutes for
$J=0$ mesons both widens the range of possibilities for extending
the standard electroweak model and, together, yields new constraints on them.

\vskip 1cm 
\begin{em}
\underline {Acknowledgements}: It is a pleasure to thank M. Banner and M.B.
Gavela for stimulating and enlightening discussions.
\end{em}
\listoffigures
\medskip
\begin{em}
Fig. 1: decay $P \rar P_1 P_2 P_3$ in the quark picture:
                   factorizable diagrams; \l
Fig. 2: decay $P \rar P_1 P_2 P_3$ in the quark picture:
                   non-factorizable diagrams;\l
Fig. 3: decay $P \rar P_1 P_2 P_3$ in the present model for $J=0$ mesons.
\end{em}
\newpage\null
\voffset -3truecm
\begin{em}

\end{em}
\end{document}